\documentclass[aps,prd]{revtex4}

\usepackage{graphicx}
\usepackage{dcolumn}
\usepackage{bm}
\RequirePackage{newlfont}

\begin{document}


\title{Asymptotic Behavior of a Class of Expanding Gowdy Spacetimes}

\author{Beverly K. Berger}
\email{bberger@nsf.gov}
\affiliation{Physics Division, National Science Foundation, Arlington, VA  22230
USA}

\date{\today}

\begin{abstract}
A systematic asymptotic expansion is developed for the gravitational wave degrees
of freedom of a class of expanding, vacuum Gowdy cosmological spacetimes. In the
wave map description of these models, the evolution of the gravitational wave
amplitudes defines an orbit in the target space. The circumference of this orbit
decays to zero indicating that the asymptotic spacetime is spatially homogeneous. A
prescription is given to identify the asymptotic cosmological model for the
gravitational wave degrees of freedom using the asymptotic point in the target
space. The remaining metric function of the asymptotic cosmological model,
found by solving the constraints, is determined by an effective energy
of the gravitational waves rather than from the asymptotic point in the target
space. 
\end{abstract}

\pacs{98.80.Dr, 04.20.J}
\maketitle
\section{\label{introduction}Introduction}
Current understanding of cosmology relies heavily on the assumption that the
large-scale dynamics of the observable universe may be described by an
averaged matter distribution evolving in a spatially homogeneous (and isotropic)
background. Yet it is clear that the universe is characterized by spatial
inhomogeneities that may be large and might be dynamically relevant. Some studies
have been made to provide a rigorous basis for the validity of the standard
cosmological approach but the question is not settled
\cite{kasai93,tomita94,deruelle95,carfora95,futamase96,stoeger99,buchert00}.  

The Gowdy cosmological spacetimes \cite{gowdy71,berger74} have proven to be valuable
theoretical laboratories to explore the behavior of spatially inhomogeneous
cosmologies. For example, polarized and generic Gowdy models on $T^3 \times R$
provided the first detailed demonstration of the Belinskii et al (BKL)
\cite{belinskii71b} claim that, in the collapse direction, each spatial point of a
generic cosmological spacetime evolves toward the singularity as a separate
universe \cite{isenberg90,berger93,berger97b}.

Polarized Gowdy models have also been studied in the expanding direction, both for
$T^3$ \cite{berger74} and $R^3$ \cite{centrella82} spatial topology, where it was
argued that the asymptotic behavior could be understood as a stiff fluid of
gravitational waves propagating in a spatially homogeneous background
spacetime. Related numerical
studies of fluids in plane symmetric expanding cosmologies have also been discussed
\cite{anninos98,anninos99}. In this paper, we will argue that this type of
asymptotic picture, previously obtained for the polarized case, persists for
a large class of generic Gowdy models. We shall also demonstrate how the background
cosmology may be extracted from a numerical simulation of these spacetimes.

Ringstr{\"{o}}m has recently shown \cite{ringstrom01} that expanding Gowdy
spacetimes fall into two classes with qualitatively different behaviors. Only one
of these classes is considered here.

\section{\label{model} Gowdy model on $T^3 \times R$}
The vacuum Gowdy cosmology on $T^3 \times R$ is described by the metric
\cite{gowdy71,berger93,berger97b}
\begin{equation}
\label{gowdymetric}
ds^2 =
e^{\lambda/2}\,t^{-1/2}\,(-\,dt^2\,+\,d\theta^2)\,+\,t\,\left[e^P\,(dx\,+\,Q\,
dy)^2\,+\,e^{-P}\,dy^2 \right]
\end{equation}
where $0 < t < \infty$, $\theta \,\in \, [0,2\pi]$, $\{x,y\}\, \in \, T^2$, and
the gravitational wave amplitudes $P$, $Q$ and the background variable $\lambda$
depend only on $\theta$ and $t$. If one assumes that $P$ and $Q$ are small, it
is easy to see that $P$ is related to the amplitude of the $+$ polarization and $Q$
to the $\times$ polarization of gravitational waves.  (See \cite{berger74,berger93}
for discussion of the interpretation of these variables.) Einstein's equations
become (with the notation that
$z,_t =
\partial z /
\partial t$)
\begin{equation}
\label{peq}
P,_{tt} \,+\, {1 \over t} P,_t \,-\,P,_{\theta \theta} \,+\, e^{2P}
\left(Q,_\theta^2 \,-\, Q,_t^2 \right) = 0
\end{equation}
and
\begin{equation}
\label{qeq}
Q,_{tt} \,+\, {1 \over t} Q,_t \,-\,Q,_{\theta \theta} \,+\, 2\,P,_t\,Q,_t
\,-\,2\,P,_\theta \,Q,_\theta = 0
\end{equation}
for the wave amplitudes and
\begin{equation}
\label{dldt}
\lambda,_t \,-\, t\,\left( P,_t^2 \,+\,e^{2P}Q,_t^2 \,+ \,P,_\theta^2 \,+\, e^{2P}
Q,_\theta^2 \right) = 0
\end{equation}
and
\begin{equation}
\label{dldx}
\lambda,_\theta \,-\, 2\,t\, (P,_t P,_\theta \,+ \,e^{2P} Q,_t Q,_\theta) = 0
\end{equation}
for the background \footnote{Eqs.~(\ref{dldt}) and (\ref{dldx}) correct a sign
error in the corresponding equations in \cite{berger93}.}. Note that
Eqs.~(\ref{dldt}) and ({\ref{dldx}) are respectively the Hamiltonian and momentum
constraints and that
$\lambda$ may be constructed after the wave equations (\ref{peq}) and (\ref{qeq})
are solved for $P$ and $Q$. The periodic boundary conditions imposed on the metric
by $T^3$ spatial topology require  
\cite{berger74}
\begin{equation}
\label{lint}
\oint \, \lambda,_\theta \, d\theta = 0
\end{equation}
which may be understood as a restriction on $P$ and $Q$ through
Eq.~(\ref{dldx}). This condition must be imposed on the initial data and is
preserved by the evolution.

Numerical simulation is straightforward since the spiky features which are
prominent in collapse \cite{berger93,berger97b} do not develop in the expansion
direction. However, in contrast to the dominance of local behavior in collapsing
models, the global behavior is dynamically important so that, at any $t$, the
spatial waveforms must be obtained accurately. 

If we specialize to the case that $P$, $Q$, and $\lambda$ depend on $t$ only,
Einstein's equations reduce to the spatially homogeneous equations (where overdot
$= d/dt$)
\cite{berger93}
\begin{equation}
\label{Pvtd}
\ddot P \,+\, {1 \over t} \dot P \,- \, e^{2P} \dot Q^2 = 0,
\end{equation}
\begin{equation}
\label{Qvtd}
\ddot Q \,+\, {1 \over t} \dot Q \,+\, 2\,\dot P \dot Q = 0,
\end{equation}
and
\begin{equation}
\label{lambdavtd}
\dot \lambda \,-\, t\,\left(\dot P^2 \,+\,e^{2P} \dot Q^2 \right) = 0
\end{equation}
with the known spatially homogeneous solution \cite{berger97c} 
\begin{equation}
\label{p0soln}
P = \ln |\mu|\, + \,v \ln \left( {t \over {t_0}} \right) \,+\,
\ln \left[ 1\,+\,\left( {t \over {t_0}} \right)^{-2v} \right],
\end{equation}
\begin{equation}
\label{q0soln}
Q = \xi \,+\, {1 \over {\mu \left[1 \,+\, \left({t \over
{t_0}} \right)^{-2v} \right]}},
\end{equation}
and
\begin{equation}
\label{l0soln}
\lambda = \lambda_0 + v^2 \,\ln \, t
\end{equation}
where $\mu$, $\xi$, $v$, $t_0$, and $\lambda_0$ are constants. For convenience,
consider only the case
$v > 0$ as $t \to \infty$. The metric becomes (where some constants have been
absorbed and with $Q \to c$ as $t \to \infty$)
\begin{equation}
\label{kasnermetric}
ds^2 = t^{(v^2-1)/2}(-dt^2\,+\,d\theta^2)\,+\,t^{1+v}(dx\,+\,cdy)^2
\,+\,t^{1-v}dy^2 .
\end{equation}
The contribution due to $Q$ (i.e. the constant $c$) rotates the coordinate axes with
respect to two of the Killing directions. In terms of comoving proper time, $T$,
defined by
\begin{equation}
\label{propertime}
T = t^{(v^2+3)/4}
\end{equation}
(where again some constants have been absorbed), the metric (\ref{kasnermetric})
becomes
\begin{equation}
\label{kasnermetrica}
ds^2 = -dT^2 \,+\,T^{2(v^2-1)/(v^2+3)} d\theta^2 \,+\, T^{4(1+v)/(v^2+3)}(dx\,+\, c
dy)^2 \,+\,T^{4(1-v)/(v^2+3)}dy^2.
\end{equation}
The exponents of the scale factors $T^{k_i}$ are clearly seen to satisfy the Kasner
relations $\textstyle \sum_{k=1}^3 k_i = 1 = \textstyle \sum_{k=1}^3 k_i^2$. 
We shall see later that the presence of spatial inhomogeneities
(gravitational waves) will modify this Kasner background spacetime.

As is well-known (see e.g.
\cite{berger93}), the Gowdy wave equations (\ref{peq}) and (\ref{qeq})
may be regarded as wave maps with the target space described by the metric
\begin{equation}
\label{targetmetric}
d{\cal S}^2 = dP^2 \,+\,e^{2P} dQ^2.
\end{equation}
A spatially homogeneous solution $\{P,Q\}(t)$ is represented by a point in the
target space while a spatially inhomogeneous solution $\{P,Q\}(t,\theta)$ yields a
closed curve in the target space. If, asymptotically as $t \to \infty$, a
spatially inhomogeneous solution approaches a spatially homogeneous one, the
circumference of the curve will asymptotically approach zero and the asymptotic
point thus obtained will identify the asymptotic spatially homogeneous solution. The
symmetries of the target space yield three constants of the motion for the spatially
homogeneous equations (\ref{Pvtd}), (\ref{Qvtd}):
\begin{equation}
\label{Atwiddle}
\tilde A = \pi_Q\,Q \,-\,\pi_P,
\end{equation}
\begin{equation}
\label{Btwiddle}
\tilde B = \pi_Q,
\end{equation}
\begin{equation}
\label{Ctwiddle}
\tilde C = \pi_Q\, \left( e^{-2P}\,-\,Q^2 \right)\,+\,2Q\,\pi_P,
\end{equation}
where $\pi_P = -tP,_t$ and $\pi_Q = -t\,e^{2P}Q,_t$ are the momenta canonically
conjugate to $P$ and $Q$. These constants may be related to the constants of the
spatially homogeneous solution (\ref{p0soln}) and (\ref{q0soln}) through
\begin{equation}
\label{vtdconstants}
v = \sqrt{\tilde A^2\,+\, \tilde B \tilde C}\ , \quad \mu = -{{\tilde B} \over
{2v}}\ ,
\quad \xi = - {{\tilde A \,+\,v} \over {2v\mu}}\ .
\end{equation}

Although $\tilde A$, $\tilde B$, and $\tilde C$ are not constants of the motion for
the wave equations (\ref{peq}) and (\ref{qeq}), their spatial averages $A$, $B$,
and $C$ are constants of the motion. The dynamics of spatially inhomogeneous Gowdy
models falls into two classes depending on the sign of $A^2 + BC$
\cite{ringstrom01}. Here we consider only $A^2 + BC > 0$, the only case allowed
for solutions to the spatially homogeneous equations (\ref{Pvtd}) and
(\ref{Qvtd}) and the case assumed in the definition of $v$ in
Eq.\ (\ref{vtdconstants}). We shall show later that knowledge of $A$, $B$, and $C$
allows the asymptotic point in the target space associated with the asymptotic
spatially homogeneous solution to be identified.

\section{\label{perturbation}Perturbation Analysis}
To understand the asymptotic ($t \to \infty$) behavior of this class of expanding
Gowdy models and the dynamical role of the gravitational waves assume that $P$ and
$Q$ may be expressed as asymptotic expansions
\begin{equation}
\label{pseries}
P = p_0(t) \,+\, \varepsilon \,p_1(\theta,t) \,+\, \varepsilon^2 \,p_2(\theta,t)
\,+\, \ldots,
\end{equation}
\begin{equation}
\label{qseries}
Q =  \varepsilon \, q_1(\theta,t) \,+\, \varepsilon^2 \, q_2(\theta,t)
\,+\, \ldots
\end{equation}
where $\varepsilon$ is a small parameter and the order $\varepsilon^0$ term in $Q$
is set equal to zero.. These asymptotic forms are substituted into Eqs.~(\ref{peq})
and (\ref{qeq}) and terms collected according to their order in powers of
$\varepsilon$. 

In the development of this perturbation series, it is convenient to insist as we
have done here that the order $\varepsilon^0$ solution be polarized---i.e., we
require
$q_0(t) = 0 =
\dot q_0(t)$. Then the order $\varepsilon^0$ equations are just the polarized
version of Eqs.~(\ref{Pvtd}) and (\ref{Qvtd}) which reduce to 
\begin{equation}
\label{p0eq}
\ddot p_0 \,+\, {1 \over t} \dot p_0 = 0
\end{equation}
with the general solution
\begin{equation}
\label{p0solnpol}
p_0 = \alpha \,+\, v \ln t
\end{equation}
where $\alpha$ and $v$ are constants. This restriction on the order $\varepsilon^0$
solution is not a loss of generality within this class of solutions since any
solution to Eqs.~(\ref{Pvtd}) and (\ref{Qvtd}) with $A^2 + BC > 0$ may be obtained
from a polarized solution through an $SL(2,R)$ transformation \footnote{The
polarized limit of the solution (\ref{p0soln}), (\ref{q0soln}) (for $v > 0$) is
$t_0 \to 0$, $\mu
\to (t_0)^v$, and $\xi \to 1/\mu$. A different parametrization of the homogeneous
solution (see \cite{berger97b}) can be found with a more straightforward polarized
limit. The parametrization chosen in this paper is more convenient for
establishment of the connection to $\tilde A$, $\tilde B$, and $\tilde C$ in
Eq.~(\ref{vtdconstants}).}. (See, for example, \cite{berger00}.)

To order $\varepsilon^1$, Eqs.~(\ref{peq}) and (\ref{qeq})
become
\begin{equation}
\label{p1eq}
p_1,_{tt} \,+\, {1 \over t}\, p_1,_t \,-\, p_1,_{\theta \theta} \equiv {\cal L}_P \,
p_1 =0,
\end{equation}
\begin{equation}
\label{q1eq}
q_1,_{tt} \,+\, \left({1 \over t} \,+\, {{2v} \over t} \right) q_1,_t \,-\,
q_1,_{\theta
\theta}
\equiv {\cal L}_Q
\, q_1 =0.
\end{equation}
(Note that the second term within the parenthesis in Eq.~(\ref{q1eq}) is $2 \dot
p_0$.) The general solutions to these linear equations are (with arbitrary
constants chosen to eliminate the spatially homogeneous modes of $p_1$ and $q_1$)
\begin{equation}
\label{p1soln}
p_1 = \sum_{n=1}^N \, b_n \, {\cal Z}_0 (nt) \, \cos (n\theta + \phi_n),
\end{equation}
\begin{equation}
\label{q1soln}
q_1 = t^{-v} \, \sum_{n=1}^N \, d_n \, {\cal Z}_{\pm v} (nt) \, \cos (n\theta +
\psi_n)
\end{equation}
where $n = 1,\,2,\,\ldots, N$, $a_n$, $b_n$, $\phi_n$, and $\psi_n$ are constants
with ${\cal Z}_\nu(x)$ any Bessel function of order $\nu$ \footnote{To avoid the
complication of the convergence of infinite sums, we shall restrict our
consideration to finite sums $N < \infty$.}.  We note that
the regular and irregular Bessel functions, $J_\nu(z)$ and
$Y_\nu(z)$ respectively, have the following large-argument asymptotic expansions
\cite{abramowitz65}:
\begin{equation}
\label{jnu}
J_\nu(z) = \sqrt{{2 \over {\pi z}}} \left[\cos(z-{\pi \over 2} \nu - {\pi
\over 4} ) - {{4 \nu^2 -1} \over {8z}} \sin(z-{\pi \over 2} \nu - {\pi
\over 4} )  + \ldots \right]\,, 
\end{equation}
\begin{equation}
\label{ynu}
Y_\nu(z) = \sqrt{{2 \over {\pi z}}} \left[\sin(z-{\pi \over 2} \nu - {\pi
\over 4} ) + {{4 \nu^2 -1} \over {8z}} \cos(z-{\pi \over 2} \nu - {\pi
\over 4} )  + \ldots \right]\,, 
\end{equation}
\begin{equation}
\label{jnuprime}
J'_\nu(z) = \sqrt{{2 \over {\pi z}}} \left[-\sin(z-{\pi \over 2} \nu -
{\pi
\over 4} ) - {{4 \nu^2 +3} \over {8z}} \cos(z-{\pi \over 2} \nu - {\pi
\over 4} )  + \ldots \right]\,, 
\end{equation}
\begin{equation}
\label{ynuprime}
Y'_\nu(z) = \sqrt{{2 \over {\pi z}}} \left[\cos(z-{\pi \over 2} \nu - {\pi
\over 4} ) - {{4 \nu^2 +3} \over {8z}} \sin(z-{\pi \over 2} \nu - {\pi
\over 4} )  + \ldots \right]\,, 
\end{equation}
where a prime indicates the derivative with respect to the argument.
These asymptotic forms for the Bessel functions indicate that (1) an arbitrary
superposition of Bessel functions of order $\nu$ and argument $nt$ can be
represented asymptotically as 
\begin{equation}
\label{bessel}
{\cal Z}_\nu(nt) \to \xi_n \sqrt{{2} \over {\pi nt}} \cos ( nt  + \varphi_n)
\end{equation}
where the constant $\xi_n$ includes any normalization and $\varphi_n$ is an
arbitrary phase; (2) as $t \to
\infty$, the leading terms in Eqs.~(\ref{p1soln}) and (\ref{q1soln}) of the form
${\cal Z}_\nu(nt) \cos (n\theta + \delta_n)$ have the property that the leading
terms of the derivatives with respect to $t$ (or, trivially, $\theta$) decay as
$1/\sqrt{t}$---i.e., to leading order in inverse powers of $t$, the derivative acts
only on the trigonometric functions.

To order $\varepsilon^2$, Eqs.~(\ref{peq}) and (\ref{qeq}) become
\begin{equation}
\label{p2eq}
{\cal L}_P \, p_2 = e^{2p_0} \left(q_1,_t^2 \,-\,q_1,_\theta^2 \right),
\end{equation}
\begin{equation}
\label{q2eq}
{\cal L}_Q \, q_2 = 2\,p_1,_\theta \,q_1,_\theta \,-\, 2\,p_1,_t\,q_1,_t.
\end{equation}
The order $\varepsilon^2$ solutions may be expressed as Fourier series (chosen, for
convenience, to be the sum of a finite number of terms),
\begin{equation}
\label{p2fs}
p_2(\theta,t) = p_2^{(0)}(t) \,+\, \sum_{k=1}^N p_2^{(k)}(t) \cos(k\theta + \phi_k),
\end{equation}
\begin{equation}
\label{q2fs}
q_2(\theta,t) = q_2^{(0)}(t) \,+\, \sum_{k=1}^N q_2^{(k)}(t) \cos(k\theta + \psi_k).
\end{equation}
The spatially homogeneous modes $p_2^{(0)}$, $q_2^{(0)}$ could, in principle,
seriously modify the background spacetime if they are not small compared to $p_0$
(and, in general, $q_0$). Integration over the circle of Eqs.~(\ref{p2eq}) and
(\ref{q2eq}) yields
\begin{equation}
\label{p2homog}
\ddot p_2^{(0)} \,+\, {1 \over t}\, \dot p_2^{(0)} = {1 \over {2 \pi}}\oint d \theta
\, e^{2p_0}
\left[ (q_1,_t)^2 \,-\, (q_1,_\theta)^2 \right] \equiv {\cal S}_P(t), 
\end{equation}
\begin{equation}
\label{q2homog}
\ddot q_2^{(0)} \,+\, {{1 + 2v} \over t}\, \dot q_2^{(0)} = {1 \over {\pi}} \, \oint
d
\theta
\,
\left( p_1,_\theta q_1,_\theta \,-\, p_1,_t q_1,_t \right) \equiv {\cal
S}_Q(t), 
\end{equation}
Wronskian methods \cite{arfken00} may be used to solve these equations. Note that
the Wronskian for Bessel's equation is
\begin{equation}
\label{wbessel}
W = J_\nu(z) Y'_\nu(z) - Y_\nu(z) J'_\nu(z) = {2 \over {\pi z}}\,.
\end{equation}

If a second order homogeneous ordinary differential equation (ODE) has linearly
independent solutions
$y_1(t)$ and $y_2(t)$, then the general
solution to the inhomogeneous ODE with a source term $s(t)$ is an arbitrary linear
combination of
$y_1$ and $y_2$ plus the particular solution
\begin{equation}
\label{inhomogsoln}
\int^{t} s(t') dt' \left [ {{y_1(t') y_2(t) - y_2(t') y_1(t)} \over
{y_1(t') y'_2(t')- y_2(t') y'_1(t')}} \right] 
\end{equation}
where the quantity in the denominator is the Wronskian.

First apply Eq.~(\ref{inhomogsoln}) to Eq.~(\ref{p2homog}). 
Linearly independent solutions to the (mathematically) homogeneous
equation are $y_1 = 1$ and $y_2 = \ln t$ with Wronskian $W = 1/t$ while
the source $s(t)$ is just ${\cal S}_P(t)$. We find
\begin{equation}
\label{p20soln}
p_2^{(0)} = \ln t \int^t {\cal S}_P(t')\, t' dt' - \int^t \ln t' \,{\cal
S}_P(t')\, t' dt' \,. 
\end{equation}
Partial intgration of the second term cancels the first term on the right hand
side (RHS) and leaves 
\begin{equation}
\label{p20solna}
p_2^{(0)} (t) = \int^t dt'\,{1 \over {t'}} \int^{t'} {\cal S}_P(t'')\, t''dt''
\,. 
\end{equation}
Now $p_1$ and $q_1$ from Eqs.~(\ref{p1soln}) and (\ref{q1soln}) and their space and
time derivatives may be used to construct the source terms ${\cal S}_P(t)$ and
${\cal S}_Q(t)$.  Note that, from Eq.\ (\ref{q1soln}), 
\begin{equation}
\label{dq1dt}
q_1,_t = -{v \over t} q_1 \,+\, t^{-v} \sum_{n=1}^N \, d_n \dot {\cal Z}_{\pm v}
(nt) \, \cos(n\theta + \psi_n).
\end{equation}
From Eqs.\ (\ref{jnuprime}) and (\ref{ynuprime}), it is clear that, as $t \to
\infty$, the first term on the RHS may be neglected compared to the
second. Such terms arising from the time derivative of the factor $t^{-v}$ in
$q_1$ will be neglected in the following discussion.  We find that the right hand
sides of Eqs.~(\ref{p2eq}) and (\ref{q2eq}) yield the spacetime dependent source
terms
\begin{eqnarray}
\label{spthetat}
{\cal S}_P(\theta,t) &\approx& \sum_{n=1}^N \sum_{m=1}^N \,e^{2\alpha}\,\left[(d_n
d_m nm{\cal Z}'_v(nt) {\cal Z}'_v(mt) \cos(n\theta + \psi_n)
\cos(m\theta+\psi_m) \right. \nonumber \\
 &&- \left. d_n d_m mn{\cal
Z}_v(nt) {\cal Z}_c(mt) \sin(n\theta + \psi_n) \sin(m\theta+\psi_m)
\right] \, , 
\end{eqnarray}
\begin{eqnarray}
\label{sqthetat}
{\cal S}_Q(\theta,t) &\approx& -t^{-v}\sum_{n=1}^N \sum_{m=1}^N
\left[b_n d_m nm{\cal Z}'_0(nt) {\cal Z}'_v(mt) \cos(n\theta + \phi_n)
\cos(m\theta+\psi_m) \right. \nonumber \\
& &- \left. b_n d_m mn{\cal
Z}_0(nt) {\cal Z}_v(mt) \sin(n\theta + \psi_n) \sin(m\theta+\phi_m)
\right] \, . 
\end{eqnarray}
Of course, the products of sines and cosines may be written as cosines of
the sums and differences of the arguments. Integration over the circle
will eliminate the sum of argument terms. We thus obtain the time
dependent source terms
\begin{equation}
\label{sp}
{\cal S}_P(t) \approx {1 \over 2} \sum_{n=1}^N d_n^2 n^2 \,e^{2\alpha}\,\left[
({\cal Z}'_v(nt))^2 - ({\cal Z}_v(nt))^2 \right] \,, 
\end{equation}
\begin{equation}
\label{sq}
{\cal S}_Q(t) \approx - t ^{-v}{1 \over 2}\sum_{n=1}^N b_nd_n n^2
\cos(\psi_n-\phi_n)
\left[{\cal Z}'_0(nt) {\cal Z}'_v(nt) - {\cal Z}_0(nt) {\cal Z}_v(nt)
\right] \,. 
\end{equation}

Assume that the mixture of $J_\nu$ and $Y_\nu$ is accounted
for by an arbitrary phase in the argument of the trigonometric function
in the leading term of the asymptotic expansion. Thus, let ${\cal Z}_v (nt) \approx
\sqrt{2/\pi n t} \,
\sin(nt + \zeta_n)$ and ${\cal Z}'_v (nt) \approx \sqrt{2/\pi n t} \,\cos(nt
+ \zeta_n)$ so that
\begin{equation}\,
{\cal S}_P(t) \approx \sum_{n=1}^N d_n^2  {{n}\over {\pi t}}\, e^{2\alpha}
\left[
\cos^2 (nt + \zeta_n) - \sin^2(nt + \zeta_n) \right] \approx 
 \sum_{n=1}^N d_n^2  \,{{n} \over {\pi t}}\,e^{2\alpha}
\cos (2 nt + 2 \zeta_n)\,. 
\end{equation}
Substitution into Eq.~(\ref{p20solna}) yields 
\begin{equation}
\label{p20solnb}
p_2^{(0)} \approx  \sum_{n=1}^N d_n^2  \, e^{2\alpha}\,\int^t dt'{{1} \over {2\pi
t'}}
\left[ \sin(2nt' +2 \zeta_n) - \chi_n \right] 
\end{equation}
where $\chi_n$ is a constant of integration. 

Asymptotically, 
\begin{equation}
\label{asympttrig}
\int^t {{\sin(at' + b)} \over t'} dt' \approx -{{ \cos(at + b)} \over {at}}
\end{equation}
since the difference between the derivative of the RHS of Eq.~(\ref{asympttrig}) and
the integrand on the left hand side goes as $t^{-2}$. Using this argument,
\begin{equation}
\label{p20solnc}
p_2^{(0)} \approx  -\sum_{n=1}^N d_n^2 \,e^{2\alpha}\,\left[ {{1} \over {4\pi n t}}
\sin(2nt +2 \zeta_n)- {\xi_n \over {2 \pi}} \ln t \,+\, \sigma_n \right]
\end{equation}
where $\sigma_n$ is a constant.
It is clear that $p_2^{(0)}$ is indeed a perturbation of $p_0 = v \ln t$ since
an arbitrary amount of the homogeneous solution of Eq.\ (\ref{p2homog}) may be added
to eliminate the logarithmic and constant terms in Eq.~(\ref{p20solnc}).

Now solve Eq.\ (\ref{q2homog}) for $q_2^{(0)}$. Here the linearly independent
solutions are $y_1 = 1$ and $y_2 = t^{-2v}$ with the Wronskian $W = -2v\,
t^{-2v-1}$. Substitution into Eq.~(\ref{inhomogsoln}) shows that the particular
solution to Eq.~(\ref{q2homog}) is
\begin{equation}
\label{q20solna}
q_2^{(0)} (t)= - {1 \over {2v}} t^{-2v} \int^t {\cal S}_Q(t') t'^{2v+1}
dt' +{1 \over {2v}} \int^t {\cal S}_Q(t') t' dt' \,. 
\end{equation}
Taking the time derivative and then integrating shows that Eq.~(\ref{q20solna}) is
equivalent to
\begin{equation}
\label{q20solnb}
q_2^{(0)} (t) = \int^t dt' (t')^{-2v-1} \, \int^{t'} dt'' (t'')^{2v+1}\,S_Q(t'') \,.
\end{equation}
Just as before, we evaluate the asymptotic form of ${\cal S}_Q(t)$. With the
previous asymptotic form for ${\cal Z}_v$ and its derivative and 
${\cal Z}_0 (nt) \approx \sqrt{2/\pi n t}\, \sin(nt
+ \xi_n)$ and ${\cal Z}'_0 (nt) \approx \sqrt{2/\pi n t}\, \cos(nt
+ \xi_n)$, 
\begin{eqnarray}
\label{sqeval}
{\cal S}_Q(t) &\approx & \sum_{n=1}^N b_n d_n \, {{n} \over {\pi \,t^{1+v}}}
\left[ \cos(nt + \xi_n)
\cos (nt + \zeta_n) - \sin(nt + \xi_n) \sin(nt + \zeta_n) \right] \nonumber \\
&\approx &
 \sum_{n=1}^N b_n d_n {{n} \over {\pi t^{1+v}}}
\cos (2 nt + \zeta_n +\xi_n) \,.
\end{eqnarray}
After partial integration, the integral over $t''$ in (\ref{q20solnb}) yields as
the dominant term
\begin{equation}
\label{q20solnc}
q_2^{(0)} \approx -\sum_{n=1}^N b_n d_n  \, \int^t dt'\, (t')^{-v-1}
\, \sin (2 nt' + \zeta_n +\xi_n)  \,. 
\end{equation}
A multiple of the homogeneous solution has been added to eliminate constants of
integration. The same argument [Eq.~(\ref{asympttrig})] concerning asymptotics of
the integrals is valid here. Essentially, the leading term in Eq.~(\ref{q20solnc})
is obtained by integrating only the trigonomentric term. Thus we obtain
\begin{equation}
\label{q20solnd}
q_2^{(0)} \approx \sum_{n=1}^N b_n d_n {{1} \over {2n\pi {t^{v+1}}}}
\cos (2 nt + \zeta_n +\xi_n) \, .
\end{equation}
Again, this quantity is seen to be a decaying perturbation on $q_0$'s constant
value.  

The solutions $p_2^{(0)}$ and $q_2^{(0)}$ decay as $t^{-1}$ and $t^{-v-1}$
respectively---i.e. faster than
$p_1$ and $q_1$---and are perturbations of $p_0$ and $q_0$. [For a general
(unpolarized) order $\varepsilon^0$ solution, as $t \to \infty$, $p_0 \to v \ln t$
and $q_0 \to c$, a constant.] The next relevant order for the spatially homogeneous 
mode, $p_3^{(0)}$
and $q_3^{(0)}$, will have source terms containing (e.g.) $e^{p_0}
(q')_1^2 \,p_1$ that will decay as $t^{-3/2}$, with every other
aspect of the argument the same.

Now consider the spatially dependent mode amplitudes $p_2^{(k)}$ and
$q_2^{(k)}$ for $k \ge 1$. Integrate $\cos(k\theta + \psi_k) \,{\cal
S}_p(\theta,t)$ over the circle to yield the $k$th component
\begin{eqnarray}
\label{spqt}
{\cal S}_P^{(k)}(t) & \approx& \sum_{n=1}^N \left[ d_n d_{|k-n|} n |k-n| 
{\cal Z}'_v(nt) {\cal Z}'_v(|k-n|t) \right. \nonumber \\
&& \left. - d_n d_{|k-n|}
n |k-n|  {\cal Z}_v(nt) {\cal Z}_v(|k-n|t)  \right] \nonumber \\
&+&\sum_{n=1}^N \left[ d_n d_{k+n} n (k+n) 
{\cal Z}'_v(nt) {\cal Z}'_v((k+n)t)  \right. \nonumber \\
& &\left. - d_n d_{k+n}
n (k+n) {\cal Z}_v(nt) {\cal Z}_v((k+n)t) \right]
\,.
\end{eqnarray}

We again construct the particular solution $p_2^{(k)}$ using (\ref{inhomogsoln})
where
$y_1 = J_0(kt)$, $y_2 = Y_0(kt)$, and $W = 2/(\pi kt)$ . Then
\begin{equation}
\label{p2k}
p_2^{(k)} = Y_0(kt) \int^t {\cal S}_p^{(k)}(t') J_0(kt') {{\pi kt'} \over 2} dt' -
J_0(kt) \int^t {\cal S}_p^{(k)}(t') Y_0(kt') {{\pi kt'} \over 2} dt' \,. 
\end{equation}
We can regard the asymptotic expansions of $J_\nu$ and $Y_\nu$ to be those
given in (\ref{jnu}) and (\ref{ynu}) while ${\cal Z}_\nu$ and its derivative have an
arbitrary phase.

Asymptotically,
\begin{eqnarray}
\label{spqta}
{\cal S}_P^{(k)}(t) \approx && \sum_{n=1}^N \left\{ d_n d_{|k-n|}  |k-n|
{1 \over t} cos[(n+|k-n|)t + 2 \psi_n]  \right\} \nonumber \\
&+&\sum_{n=1}^N \left\{ d_n d_{k+n}  (k+n) {1 \over t} cos[(2n + k)t + 2
\psi_n]  \right\}
\,.
\end{eqnarray}
Rather than complete the evaluation of Eq.~(\ref{spqta}), notice that
power counting gives the dominant asymptotic behavior as $t^{-1}$. The same
type of asymptotic argument about the integral of a trigonometric function
times a power law that was used previously implies that $p_2^{(k)}$ should be
an oscillatory function divided by $t$ as expected. The argument for $q_2^{(k)}$
should be identical except that appropriate powers of $t^{-v}$ must be carried
along.

We note that in general, the sums and products of trigonometric functions will
yield trigonometric functions. (Any potentially troublesome constant term has
already been considered in the sources for $p_2^{(0)}$ and $q_2^{(0)}$.) As long as
there are trigonometric functions, the leading order as $t \to \infty$ of time
derivatives of
$p_2^{(k)}$ and $q_2^{(k)}$ will arise from the derivative acting only on the
trigonometric term and not on the power law. As we have previously argued, the
corresponding statement is also true for integrals over time. Thus we expect 
$p_2^{(k)}$ and $q_2^{(k)}$ to have the forms
\begin{equation}
\label{p2nq2n}
p_2^{(k)} \approx {{{\cal P}_2^{(k)}(t,\theta)} \over t} \quad , \quad q_2^{(k)}
\approx {{{\cal Q}_2^{(k)}(t,\theta)} \over {t^{v+1}}}
\end{equation}
where ${\cal P}_2^{(k)}$ and ${\cal Q}_2^{(k)}$ are oscillatory functions of
$\theta$ and $t$.

Continuing these arguments yields that at each order we obtain for $p_n$ and $q_n$ 
functions which decay faster than those of the previous order by $\sqrt{t}$ as $t
\to \infty$. This behavior arises because, in general, $p_n$ and $q_n$ for $n
\ge 2$ will be the solution to an inhomogeneous wave equation with source terms
of the form $(``q"_j)^k\,(``p"_l)^r$ where $kj+lr = n$ and $``\ "$ indicates that
space and time derivatives may act on $q_j$ and $p_l$. (Asymptotically, these
derivatives act only on the trigonometric functions so that power counting yields
the correct power law time dependence. Note that $q_n$ will always have an extra
factor of $t^{-v}$ in its power law time dependence.) The fundamental quantities in
the source terms are $p_1$ and $q_1$ (e.g.\ appearing quadratically in the source
terms for $p_2$ and $q_2$) which decay as $t^{-1/2}$. In addition, from
Eq.~(\ref{p2nq2n}), all orders $(k)$ in the Fourier series for $p_n$ and $q_n$ for
$n \ge 1$) will have the same power law dependence on $t$. Substitution of
Eq.~(\ref{p2nq2n}) in the Fourier series (\ref{p2fs}), (\ref{q2fs}) will thus
yield
\begin{equation}
\label{pasymp}
P \approx \alpha \,+\, v \ln t \,+\, \sum_{j=1}^{N} \, {{{\cal P}_j} \over {t
^{j/2}}},
\end{equation} 
\begin{equation}
\label{qasymp}
Q \approx c\,+\, \sum_{j=1}^{N} \, {{{\cal Q}_j} \over {t ^{v+j/2}}}
\end{equation} 
where 
\begin{equation}
\label{ptrig}
{\cal P}_j = \sum_k {\cal P}_j^{(k)} (t) \, \cos(k\theta + \psi_k),
\end{equation}
\begin{equation}
\label{qtrig}
{\cal Q}_j = \sum_k {\cal Q}_j^{(k)} (t) \, \cos(k\theta + \phi_k).
\end{equation}
In the rest of this paper, we shall consider evidence for the eventual dominance of
the asymptotic behavior given by Eqs.~(\ref{pasymp}) and (\ref{qasymp}). 

\section{\label{decay} Decay to a spatially homogeneous cosmology}
The asymptotic forms (\ref{pasymp}) and (\ref{qasymp}) suggest that expanding Gowdy
spacetimes approach the spatially homogeneous cosmology described by $(p_0,q_0)$,
the solutions (\ref{p0soln}) and (\ref{q0soln}) (in the general case). Recall that
these solutions are completely determined by the constants $v$, $\mu$, $\xi$, and
$t_0$ and that the first three of these are related to the constants $\tilde A$,
$\tilde B$, and
$\tilde C$ of the spatially homogeneous solution defined in
Eqs.~(\ref{Atwiddle})--(\ref{Ctwiddle}) through Eq.~(\ref{vtdconstants}).
While $\tilde A$, $\tilde B$, and $\tilde C$ are not constants of the motion for
the wave equations (\ref{peq}), (\ref{qeq}), their integrals over the circle,
\begin{equation}
\label{constants}
A = {1 \over {2\pi}} \oint \, \tilde A\,d\theta\ , \quad B = {1 \over {2\pi}}\oint
\, \tilde B\,d\theta\ ,
\quad C = {1 \over {2\pi}} \oint \, \tilde C\,d\theta\ , 
\end{equation}
are constants of the motion. These are the well-known conserved quantities
associated to the Killing symmetries of the target space of a harmonic map---in
this case the hyperbolic plane. If $P$ and $Q$ evolve to $p_0$ and
$q_0$ respectively as $t \to \infty$, then they will approach the one parameter
family of solutions such that $A$, $B$, $C$ fix $v$, $\mu$, $\xi$ through
Eq.~(\ref{vtdconstants}) with $\tilde A$ replaced by $A$, etc. The constants $A$,
$B$, and $C$ determine a point in the target space. For any solution to the
spatially homogeneous equations (\ref{Pvtd}) and (\ref{Qvtd}), $A^2 + BC > 0$.
Since the perturbation analysis considered here expands about a solution to Eqs.\
(\ref{Pvtd}) and (\ref{Qvtd}), the signature $A^2 + BC > 0$ is built in. However,
the opposite sign is allowed for spatially inhomogeneous Gowdy models and has been
shown by Ringstr{\"{o}}m \cite{ringstrom01} to yield a qualitatively different
asymptotic state.

$A$, $B$, and $C$ may be constructed perturbatively using (\ref{pseries}) and
(\ref{qseries}). If we consider such an expansion for any of the constants, say
$B = \sum_n b_n \varepsilon^n$, we can show by interchanging derivatives with
respect to
$t$ and $\varepsilon$ that each coefficient $b_n$ in $B$ must be separately
constant. However, in general, $b_n$ will have a nontrivial power law dependence
on $t$ for $n > 0$. For $b_n$ for $n > 0$ to be constant, the coefficient of the
power law must vanish to yield $B = b_0$, the value obtained from the
homogeneous solution $p_0$, $q_0$. The same argument may be used for $A$ and
$C$. Furthermore, in general, at least for the first few terms in the asymptotic
expansion, the formal time dependence may be a growing power law. We again consider
$B$ as an explicit example.
\begin{equation}
\label{beqdetail}
B \equiv \sum_{n=0}^N b_n \varepsilon^n = -{1 \over {2 \pi}} \oint d\theta \,t \,
e^{2p_0}\,
\left[ 
\dot q_0 \, \varepsilon^0 \,+\,  q_1,_t
\, \varepsilon \,+ \, (2 p_1 q_1,_t \,+\, q_2,_t) \varepsilon^2 \,+\, \ldots
\right] \,.
\end{equation}
In the perturbation analysis of the previous section, we have assumed $q_0 = 0 =
\dot q_0$. Thus the order $\varepsilon^0$ term vanishes. This yields the expected
value for polarized solutions, $B = 0$. The order $\varepsilon$ term vanishes
because the spatial average of $q_1$ is zero. Note, however, that the integrand has
the overall asymptotic time dependence of $t^{v + 1/2}$ and thus will grow for $v >
-1/2$. The order
$\varepsilon^2$ term,
$b_2$, in (\ref{beqdetail}) is 
\begin{eqnarray}
\label{b2term}
b_2 &=& {1 \over {2 \pi}} \oint d\theta \, t^{1 + 2v} \left( 2 p_1 q_1,_t \,+\,
q_2,_t \right)
\nonumber
\\
& = & t^{1 + 2v} \left( \dot q_2^{(0)} \,+\,{1 \over {2 \pi}}  \oint d\theta\, 2
\,p_1 \, q_1,_t
\right).
\end{eqnarray}
Unless $v < -1/2$, the overall $t$ dependence indicates a growing term. However,
from (\ref{q20solnb})
\begin{equation}
\label{qdoteq}
\dot q_2^{(0)} =  (t)^{-2v-1} \, \int^{t} dt' (t')^{2v+1}\,{\cal S}_Q(t') \,.
\end{equation}
Substition of the source term written as the integral over the circle of the right
hand side of Eq.~(\ref{q2eq}) yields
\begin{equation}
\label{q20dot}
\dot q_2^{(0)} = {2 \over {t^{1+2v}}}\, \int^t \,dt' \,{1 \over {2 \pi}} \oint
d\theta \, (t')^{1+2v} \left(p_1,_\theta \, q_1,_\theta \,-\, p_1,_{t'} \,
q_1,_{t'} \right).
\end{equation}
Partially integrating the first term on the RHS with respect to $\theta$ and second
with respect to $t'$ yields
\begin{equation}
\label{q2dota}
\dot q_2^{(0)} = - 2 {1 \over {2 \pi}} \oint d\theta \, p_1 \, q_1,_t \,-\,{2  \over
{t^{1+2v}}}
\int^t
\, dt'\, {1 \over {2 \pi}}
\oint d\theta \, p_1 \, \left\{ (t')^{1+2v} \, q_1,_{\theta \theta} \,-\, {d \over
{dt'}} \left[ (t')^{2v+1} q_1,_{t'} \right] \right\} \,.
\end{equation}
The expression inside the braces is just the wave equation (\ref{q1eq}) for $q_1$
and thus vanishes to give
\begin{equation}
\label{b2coeff}
\dot q_2^{(0)} \,+\, \oint d\theta \,2 \,p_1 \,q_1,_t =0.
\end{equation}
Note that the values of $A$,
$B$, and $C$ are determined from the order $\varepsilon^0$ solution. The
order $\varepsilon^n$ terms for $n \ge 1$ vanish due to the wave equations, even
though the integrands $\tilde A$, $\tilde B$, and $\tilde C$ may grow in $t$.

To identify completely the background spacetime also requires the construction of
$\lambda$ through Eq.~(\ref{dldt}). We now show that the presence of
gravitational waves changes the time dependence of the spatially homogeneous mode of
$\lambda$ from that of the spatially homogeneous solution (\ref{l0soln}) (see
\cite{berger74}). Recall that if, asymptotically, 
\begin{equation}
\label{plimforlambda}
P \approx v \ln t\,+\, {{{\cal P}_1} \over {\sqrt{t}}},
\end{equation}
\begin{equation}
\label{qlimforlambda}
Q \approx c \,+\, {{{\cal Q}_1} \over {{t^{1/2+v}}}},
\end{equation}
then the asymptotic forms of their time derivatives are dominated as $t \to \infty$
by
\begin{equation}
\label{pdotlim}
P,_t \approx {{{\cal P}_1,_t} \over {\sqrt{t}}} \quad , \quad Q,_t \approx {{{\cal
Q}_1,_t} \over {t^{1/2+v}}} 
\end{equation}
rather than by $\dot p_0 = v/t$ where ${\cal P}_1,_t$ and ${\cal Q}_1,_t$ are
oscillatory in time. Similarly, the dominant terms in the spatial derivatives are 
\begin{equation}
\label{pthetalim}
P,_\theta \approx {{{\cal P}_1,_\theta}
\over {\sqrt{t}}} \quad , \quad Q,_\theta \approx {{{\cal Q}_1,_\theta} \over
{t^{1/2+v}}}\,.
\end{equation}
Eq.~(\ref{dldt}) then becomes 
\begin{equation}
\label{lamdotlim}
\lambda,_t \approx {\cal P}_1^2,_t
\,+\, {\cal Q}_1^2,_t\,+\, {\cal
P}_1^2,_\theta\,+\, {\cal Q}_1^2,_\theta.
\end{equation}
To compute $\bar \lambda$, the spatial average of $\lambda$, the bounded,
oscillatory functions on the RHS of Eq.~(\ref{lamdotlim}) may be replaced by a
constant average value,
$\zeta$, plus fluctuations to yield
\begin{equation}
\label{finallamlim}
\bar \lambda \approx \zeta t 
\end{equation}
in contrast to the logarithmic dependence on $t$ in Eq.\ (\ref{l0soln}).
Note that $\zeta$ depends on the details of the first order fluctuations rather
than on the constants of the spatially homogeneous solution.

Given Eq.~(\ref{finallamlim}), the metric becomes, as $t \to \infty$,
\begin{equation}
\label{metriclim}
ds^2 = e^{\zeta t /2} t^{-1/2} \left(-\,dt^2\,+\,d\theta^2 \right) \,+\, t^{1+v}
\left(dx \,+\,cdy \right)^2 \,+\, t^{1-v} dy^2.
\end{equation}
We again must transform to comoving proper time defined by
\begin{equation}
\label{limcomoving}
dT = e^{\bar \lambda /2 -1/4\,\ln t}\,dt.
\end{equation}
An approximate form of the metric (\ref{metriclim}) may be found. As $t \to
\infty$, expressions of interest may be expanded in the small quantity $(\ln t) /
t$. Thus Eq.\ (\ref{limcomoving}) becomes
\begin{equation}
\label{ttoT}
dT \approx e^{\zeta t/4}\,\left( 1 - {1 \over {\zeta t}} \ln t \right)\,dt\,.
\end{equation}
Keeping only the dominant term yields
\begin{equation}
\label{Tdef}
T \approx {4 \over {\zeta}} e^{\zeta t/4}
\end{equation}
or, neglecting some constants,
\begin{equation}
\label{tofT}
t \approx {4 \over \zeta} \ln T.
\end{equation}
This, in turn, gives the approximate metric (again absorbing constants as needed
into redefinitions of the spatial variables)
\begin{equation}
\label{approxmetric}
ds^2 \approx - dT^2 \,+\, T^2 d \tilde \theta^2 \,+\, (\ln T)^{1+v} (d \tilde x
\,+\,
\tilde c  d \tilde y)^2 \,+\,(\ln T)^{1-v} d \tilde y^2
\end{equation}
where the {\it tilde} indicates rescaling and we have used $T^2 \approx
(16/\zeta^2)\,e^{\zeta t/2}\,t^{-1/2}$. Now consider this approximate metric at
some time $T_i$ such that $\ln t(T_i) << t(T_i)$. Then, for $T > T_i$, $T = T_i
\,+\, (T - T_i)
\equiv T_i + \Delta T$ where we shall assume that $ \Delta T << T_i$. Then $\ln T
\approx \ln T_i + \Delta T/T_i$. As long as $\Delta T/T_i << \ln T_i$, the
approximate metric (\ref{approxmetric}) becomes
\begin{equation}
\label{approxmetric2}
ds^2 \approx \,-\, dT^2 \,+\, T^2\,d \tilde \theta^2 \,+\, (\ln T_i)^{1+v} (d
\tilde x + \tilde c \,d \tilde y)^2 \,+\, (\ln T_i)^{1-v} d \tilde y^2.
\end{equation}
If the spatial variables (and $\tilde c$) are defined once again to absorb powers
of $\ln T_i$, the approximate metric (\ref{approxmetric2}) may be identified as the
cosmological sector of flat spacetime in Rindler coordinates. A more detailed
discussion of the identification of the limiting metric is given for the polarized
case in \cite{berger74}.

In the following section, we shall demonstrate numerically the decay of the Gowdy
solution to this background spacetime. First we shall consider two quantities which
require the presence of the spatially {\it inhomogeneous} modes of $P$ and $Q$. Both
quantities may be understood by starting from the Hamiltonian density
\begin{equation}
\label{gowdyh}
{\cal H} = {1 \over {2t}} \left( \pi_P^2 \,+\,e^{-2P} \pi_Q^2 \right) \,+\, {t
\over 2} \left(P,_\theta^2 \,+\,e^{2P}Q,_\theta^2 \right)
\end{equation}
whose variation yields Eqs.~(\ref{peq}) and (\ref{qeq}). The quantity
\begin{equation}
\label{circ}
{\cal C} = \oint \,d\theta \left(P,_\theta^2 \,+\, e^{2P}Q,_\theta^2 \right)^{1/2}
\end{equation}
is the orbit circumference in the target space. If ${\cal C} \to 0$ as $t \to
\infty$, the solution approaches a spatially homogeneous one. Substitution of the
asymptotic forms (\ref{pasymp}), (\ref{qasymp}) into (\ref{circ}) shows that we
expect
\begin{equation}
\label{circlim}
{\cal C} \to {{\rho(t)} \over {\sqrt{t}}}
\end{equation}
where $\rho(t)$ is a bounded oscillatory function which is, asymptotically, the
spatially homogeneous mode of $({\cal P}_1,_\theta^2\,+\,{\cal Q}_1,_\theta^2
)^{1/2}$.
Another function expected to decay to zero if the asymptotic spacetime is spatially
homogeneous is the generalized force
\begin{equation}
\label{force}
{\cal F} = {1 \over t} \oint d\theta \left( \nabla {\cal H} \cdot \nabla {\cal H}
\right)^{1/2}
\end{equation}
where $\nabla = (\delta/\delta P, e^{-P} \delta / \delta Q)$ is the gradient in the
target space. We find
\begin{equation}
\label{forceeval}
{\cal F} = \oint d\theta \left[ \left( -P,_{\theta \theta} \,+\, e^{2P}Q,_\theta^2
\right)^2
\,+\, e^{-2P} \left( e^{2P} Q,_{\theta \theta} \,+\, 2e^{2P}P,_\theta Q,_\theta
\right)^2 \right]^{1/2}
\end{equation}
which has the asymptotic form
\begin{equation}
\label{forcelim}
{\cal F} \to {{\psi(t)} \over {\sqrt{t}}}
\end{equation}
with $\psi(t)$ a bounded oscillatory function.

Assuming that ${\cal C}$ and ${\cal F}$ decay, the resultant asymptotic point in
the target space is one of the spatially homogeneous solutions characterized by a particular
set of
$v$, $\mu$, $\xi$ obtained from $A$, $B$, $C$ through (\ref{vtdconstants}). The
remaining parameter
$t_0$ represents a rescaling of the time $t$. It may be determined by fitting data
obtained from numerical simulations of the full Einstein equations for the model.

\section{\label{numerical}Numerical simulations}
We use numerical simulation to provide evidence in support of our previous analysis
of the asymptotic behavior of the Gowdy spacetimes. As pointed out previously, the
lack of spiky features in the expanding spacetimes means that almost any numerical
technique can evolve the spacetime. Here we use an iterative
Cranck-Nicholson algorithm \cite{garfinkle99}. Many of our conclusions will be
based on the spatially homogeneous mode (i.e. the average over the circle) of
various quantities. Since most of these quantities are oscillatory in both space
and time with large amplitude, accurate spatial waveforms are needed to generate
the correct average. 

The quality of the numerics in the simulation may be found by computing the
conserved quantities $A$, $B$, and $C$. These are shown for representative initial
data in Fig.~\ref{fig1}. Note that $\tilde A$, $\tilde B$, and $\tilde C$ have been
singled out from among all combinations of constants of the motion of the spatially
homogeneous solution because their time derivatives are also total derivatives with
respect to $\theta$. While $v$, $\mu$, and $\xi$ are constants of the spatially
homogeneous solution, their integrals over the circle in the general case are not
constants in time. Fig.~\ref{fig2} shows one of the conserved quantities ({\it
viz.}~$A$) on a much finer scale so that deviations from constant value are
visible. These deviations display second order convergence to zero (at least during
most of the simulation). In contrast, Fig.~\ref{fig3} shows the integrands of $A$,
$B$, and $C$ at a representative spatial point. These oscillate wildly in time. The
observed constancy of $A$, $B$, and $C$ indicates that the simulation is reliable.

For a reliable simulation it is then possible to compute ${\cal C}$ and ${\cal F}$
using Eqs.~(\ref{circ}) and (\ref{forceeval}) respectively. These are shown in
Fig.~\ref{fig4} along with power law fits to the data. A closeup of the oscillatory
functions $\rho(t)$ and $\psi(t)$ (see Fig.~\ref{fig4}b) argues that the larger
discrepancy in ${\cal F}$ between the expected $t^{-1/2}$ and the observed decay is
due to the greater difficulty in determining the centroid of the pattern. Similar
decay as $t ^{-1/2}$ is shown for other representative initial data in
Fig.~\ref{fig5}.

If these solutions indeed decay to a spatially homogeneous background spacetime, it
now becomes possible to identify the precise asymptotic behavior of $P$ and $Q$.
Since $P$ and $Q$ are periodic functions on $[0,2\pi]$, it is always possible to
compute the spatially homogeneous mode of the appropriate Fourier series. This mode
for $P$ and $Q$ (say $\bar p$ and $\bar q$) contains the
$\varepsilon^0$ solution as well as the spatially homogeneous modes of the order
$\varepsilon^n$ solutions for $n \ge 2$. However, if the asymptotic forms given in
Eqs.~(\ref{pasymp}) and (\ref{qasymp}) indeed describe the solution, we require (as
argued in Section
\ref{perturbation}) that the higher order contributions to $\bar p$ and $\bar
q$ should be oscillatory (and decaying) and should oscillate about the spatially 
homogeneous solutions $p_0$ and $q_0$ respectively. Evidence for such behavior is
the following: The constants
$A$, $B$,
$C$ may be computed from the initial data and used to construct $\mu$, $\xi$, and
$v$. This yields
$p_0$ and
$q_0$ as functions of $t/t_0$ where $t_0$ is as yet unknown. Fig.~\ref{fig6} shows
that if $t_0$ is chosen to yield a best fit to (say) $\bar p$ late in the
simulation, it also fits the correct average time dependence of $\bar p$ and yields
the correct asymptotic behavior for $\bar q$. The fact that we can fit two
functions with one adjustable parameter provides strong support that the computed
$\mu$, $\xi$, $v$, and the fit $t_0$ define the correct asymptotic $P$ and $Q$.

Note that the time averaged values of $\bar p$ and $\bar q$ depart from $p_0$ and
$q_0$ early in the simulation. This means that, although $\mu$, $v$, and $\xi$ are
fixed by the initial data, $t_0$ does not have the value one would calculate from
the initial data. This means that nonlinear interactions which are important early
in the simulation influence (i.e.~renormalize) the value of $t_0$ and thus the
background spacetime. 

Also note that it is possible to evaluate $t_0$ without fitting by constructing
$\tilde p$ and $\tilde q$ defined as the solutions (\ref{p0soln}) and (\ref{q0soln})
where $\mu$, $\xi$, and $v$ are those found from the initial data while $t_0$ is
evaluated from
$\bar p$, $\bar q$ at that time step. This procedure yields $\tilde t_0(t)$ which
converges to $t_0$ as $t \to \infty$. Fig.~\ref{fig7} shows 
$\tilde p$, $\tilde q$, and $\tilde t_0$ compared to $p_0$, $q_0$, and $t_0$.

Fig.~\ref{fig8} shows the spatial average of $\lambda$ vs. $t$ for a representative
simulation. The time dependence is clearly linear (and definitely {\it not}
logarithmic). Also shown is the spatial average of the energy-like terms, ${\cal E}$
in Eq.~(\ref{dldt}), which are the ``source'' for $\lambda$. Detailed examination
shows that the fluctuations in the source terms decay (as $1/t$) to the appropriate
constant value.

The analysis of Sections \ref{perturbation} and \ref{decay} and the numerical
simulations have provided support for an asymptotic state described by
gravitational waves of decaying amplitude propagating in a spatially homogeneous
cosmology. Figures
\ref{fig9}--\ref{fig15} illustrate the mechanism yielding this asymptotic state. In
particular, the source terms ${\cal S}_P(t)$ and ${\cal S}_Q(t)$ must be driven to
small amplitude if they are not small initially. Note that the exponential
dependence on $P$ in ${\cal S}_P$ could be catastrophic if, for example, $P$ were to
depend linearly on $t$. In fact, such linear behavior of $P$ is seen very early in
the simulation in Fig.~\ref{fig12}. There, the nonlinear terms then act as
regulators (as described for the collapsing direction in
\cite{berger97b}) to suppress the linear behavior.

The consistency between the asymptotic behaviors of $P$, $Q$, and $\lambda$ found
in the numerical simulations and those expected from the analysis in Sections
\ref{perturbation} and \ref{decay} suggests that the behavior found there for the
nonlinear source terms ${\cal S}_P(t)$ and ${\cal S}_Q(t)$ on
$p_0(t)$ and $q_0(t)$ must be correct---i.e. these terms decay as $1/t$ or faster
yielding decaying perturbations $p_n^{(0)}$ and $q_n^{(0)}$ for $n \ge 2$ of $p_0$
and
$q_0$ respectively.  It is interesting to examine direct evaluation of the relevant
terms during the simulation. Figures \ref{fig9} and \ref{fig10} illustrate
respectively the component terms of ${\cal S}_P$ and ${\cal S}_Q$ (along with the
corresponding $P$ and $Q$) for the entire simulation starting from the initial data
of Fig.~\ref{fig1}. Early in the simulation, nonlinear effects dominate. The term in
${\cal S}_P$, $V_1 = \oint \, d\theta \, e^{2P}\, Q,_t^2$, in fact behaves as
$B^2\,e^{-2P}$ (i.e.\ it is dominated by the constant of motion $B$) as is seen to
be the case in Fig.~\ref{fig9} and, for more complicated initial data, in
Fig.~\ref{fig12}. Figures \ref{fig10} and
\ref{fig13} illustrate the complicated evolution of the two terms in ${\cal S}_Q$
which, in the end, are seen to decay. Figures
\ref{fig14} and \ref{fig15} show the evolution of ${\cal S}_P$ and ${\cal S}_Q$
early and late in the simulation (for the complicated initial data). Nonlinear
effects are clearly significant early in the simulation. Late in the simulation,
however, ${\cal S}_P$ and
${\cal S}_Q$ have decayed to small amplitudes with even smaller (by several orders
of magnitude) time averages (see Fig.~\ref{fig15}). This oscillatory (in time)
behavior of the source terms means that any (positive or negative) growth in
$P$ tends to produce a correction to reduce the magnitude of the growing term. Thus
the time averages of these source terms represent their actual, negligible influence
on the dynamics.

\section{\label{conclusions}Conclusions}
We have shown here that the asymptotic state of a class of expanding Gowdy models
may be described {\it quantitatively} as decaying amplitude gravitational waves
propagating in a spatially homogeneous background spacetime. Quantities such as the
orbit circumference in the target space, ${\cal C}$, and generalized force, ${\cal
F}$, are seen to decay to the background value of zero as $t^{-1/2}$ as predicted by
perturbative methods. (After these results were obtained, Ringstr\"om was able to
prove the form (\ref{circlim}) for ${\cal C}$ for ``small'' initial data
\cite{ringstrom01}.) In addition, we have provided a prescription for construction
of the asymptotic background spacetime.

Open questions concern the generality of the results discussed here. While the
asymptotic behavior (\ref{pasymp}), (\ref{qasymp}) has been found for many sets of
initial data for this class of Gowdy spacetimes, there is no guarantee that the
parameter space of such data has been fully explored. (In this context,
Ringstr\"om's result for the decay of the gravitational wave amplitudes is quite
encouraging.) One hopes that this will be settled with mathematical tools.

A more significant issue is whether or not other classes of spatially inhomogeneous
cosmological spacetimes may be described in terms of matter and gravitational
radiation with decreasing influence on a background spatially homogeneous
cosmology. Most cosmological studies assume that the present universe may be
described this way. Within the Gowdy models considered here, this type of
description arises after a nonlinear regime which has an influence on the final
state of the system. 

Investigations of more general models are in progress.  We note that, in contrast to
the collapsing case where most types of ``matter'' are dynamically unimportant,
every modification of the nature of the expanding model and its matter content
affects the dynamics. Finally, we remark that a modification of the techniques used
here have been applied to vacuum, general $T^2$ symmetric spacetimes (see
\cite{berger97g} for a description of these spacetimes). Results will be discussed
elsewhere
\cite{berger01}.

\begin{acknowledgments}
I would like to thank the Institute for Theoretical Physics at the University of
California at Santa Barbara, the Erwin Schr\"{o}dinger Institute in Vienna, the Institute for Geophysics and Planetary Physics at Lawrence Livermore National
Laboratory and the Institute for Theoretical Science at the University of Oregon
for hospitality and Vincent Moncrief and Jim Isenberg for useful discussions. This
work was supported in part by NSF Grants PHY9732629, PHY9800103, PHY00980846, and
PHY9407194. 
\end{acknowledgments}

\newpage

\begin{figure}
\includegraphics[scale = .5]{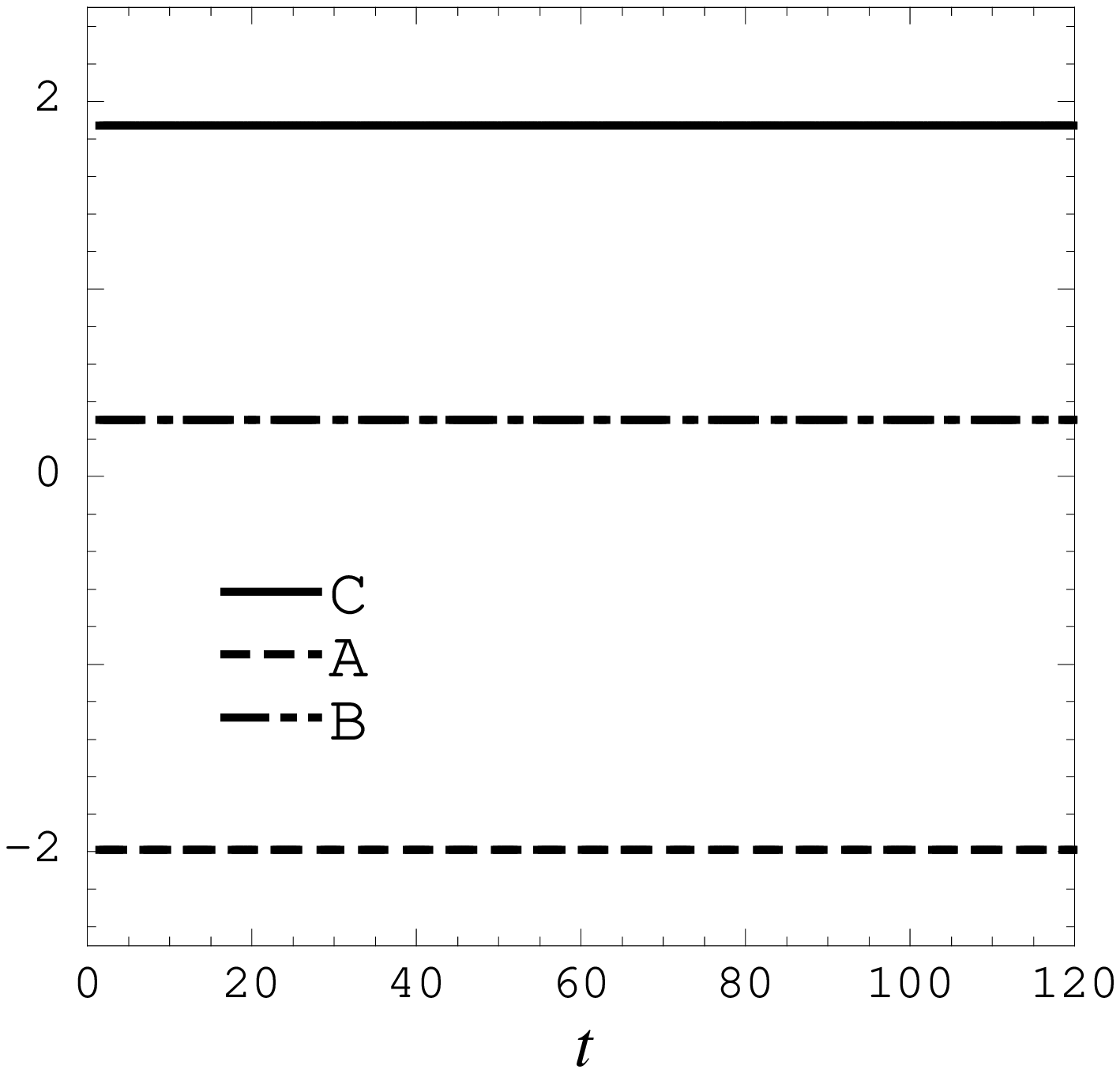}
\caption{\label{fig1} Constants of the motion $A$, $B$, and $C$ in a
representative expanding Gowdy spacetime. Unless otherwise noted, all figures are
constructed from a simulation with $P = 1$, $P,_t = 2 \,+\, 2 \cos \theta$, $Q =
\cos \theta$, $Q,_t = 0.3$, and $\lambda = 0.6 \, e^2 \, \cos \theta$ at $t = 1$.}
\end{figure}

\begin{figure}
\includegraphics[scale = .5]{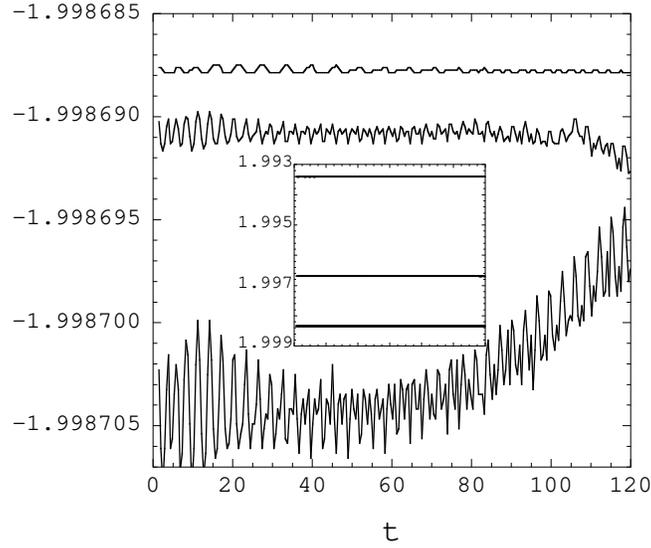}
\caption{\label{fig2} Fluctuations of $A$. On a fine numerical scale, numerical
fluctuations in $A$ (which should be strictly constant) are visible. In the main
graph, the three curves show (from top to bottom) fluctuations with spatial
resolutions of 4096, 2048, and 1024 points respectively. Offsets have been added
to the lower resolution values to allow display on the same graph. The inset shows
the mean values where the spatial resolution increases from top to bottom over the
same range in time. It is clear that numerical errors in the constant $A$ are
converging to zero and that the mean value is also converging.}
\end{figure}

\begin{figure}
\includegraphics[scale = .5]{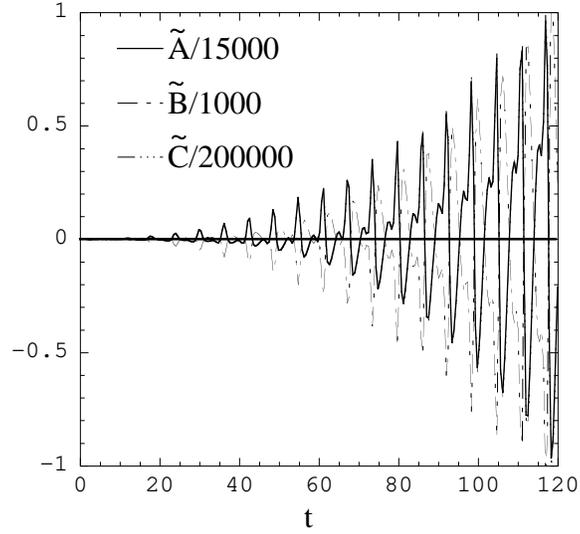}
\caption{\label{fig3} Integrands for $A$, $B$, and $C$ at a representative spatial
point. While the constants $A$, $B$, and $C$ are preserved in the numerical
simulation, the integrands $\tilde A$, $\tilde B$, and $\tilde C$ fluctuate
wildly. Here they are rescaled to allow display on a single graph. The horizontal
line represents the average values.}
\end{figure}

\begin{figure}
\includegraphics[scale = .5]{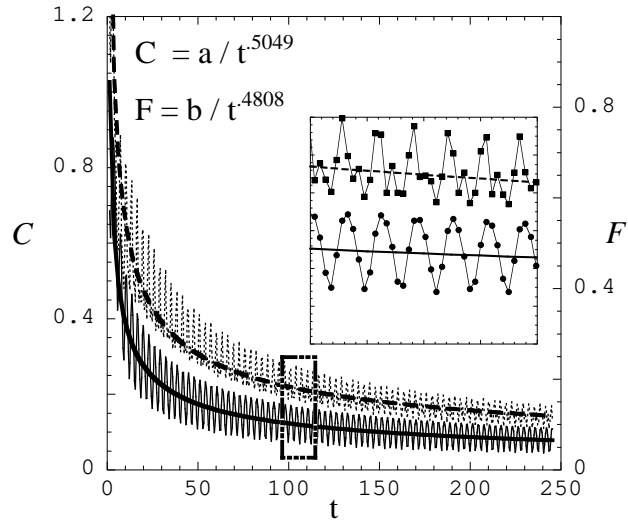}
\caption{\label{fig4} Evidence for approach to a spatially homogeneous 
background cosmology.
The circumference ${\cal C}$ and force ${\cal F}$ are shown. The dashed and solid
thick lines are power-law fits with the indicated exponents. The region inside the
rectangle is shown in the inset. The lower curve (solid line or line with circles)
represents ${\cal C}$ and the upper curve (dashed line or line with squares)
represents ${\cal F}$.}
\end{figure}

\begin{figure}
\includegraphics[scale = .5]{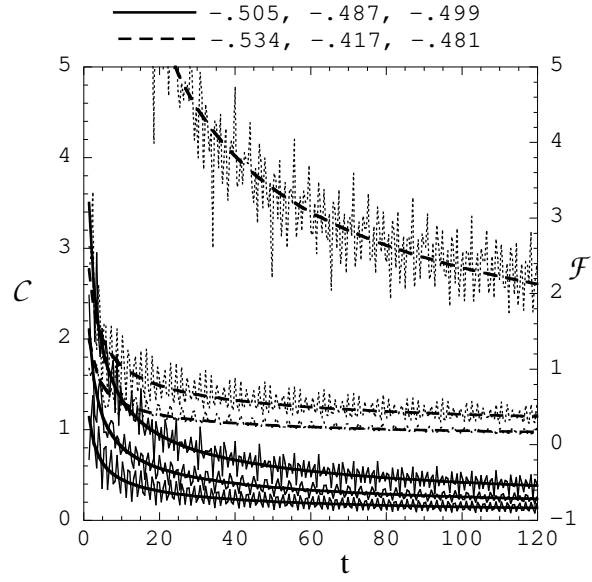}
\caption{\label{fig5} Decay of ${\cal C}$ (solid lines) and ${\cal F}$ (dashed
lines) for three different expanding Gowdy models. The respective power-law fits
are shown at the top of the graph.}
\end{figure}

\begin{figure}
\includegraphics{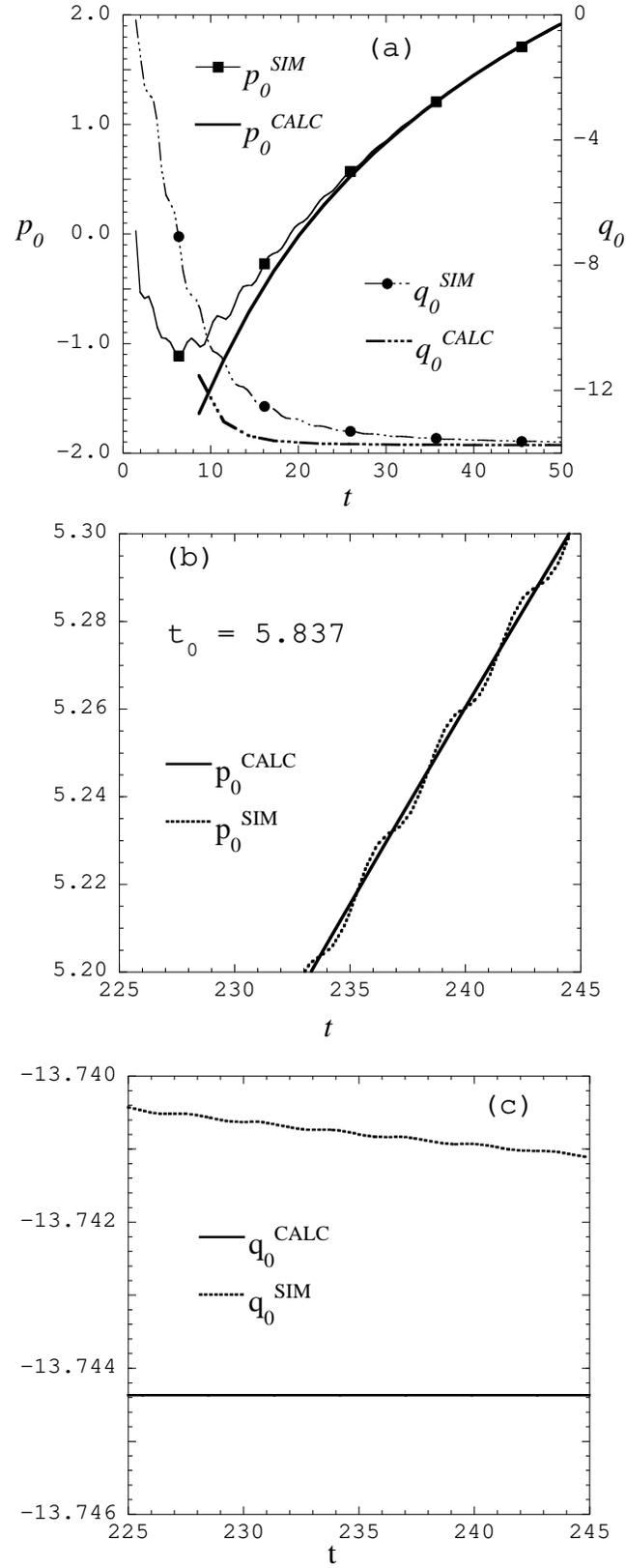}
\caption{\label{fig6} Predicting the asymptotic spatially homogeneous model. 
(a) The values of the spatially homogeneous modes, 
$\bar p$ of $P$ (solid line with squares) and $\bar q$ of
$Q$ (dashed line with circles), are compared to $p_0$ and $q_0$ calculated from
$\mu$, $\xi$, $v$, and the best fit to $t_0$. (b) Rescaling the time associated
with given values of $p_0$ and $q_0$ until $p_0$ matches the center of $\bar p$.
This occurs if $t_0 = 5.837$. (Note that the value of $t_0$ will change slightly as
the match is made for increasing $t$, eventually reaching the ``true'' value at $t
= \infty$.) (c) A comparison between $\bar q$ (dashed) and $q_0$ (solid) late in the
simulation. Note that $\bar q$ is decaying to $q_0$ and that the difference is only
a few parts in $10^4$.}
\end{figure}

\begin{figure}
\includegraphics[scale = .5]{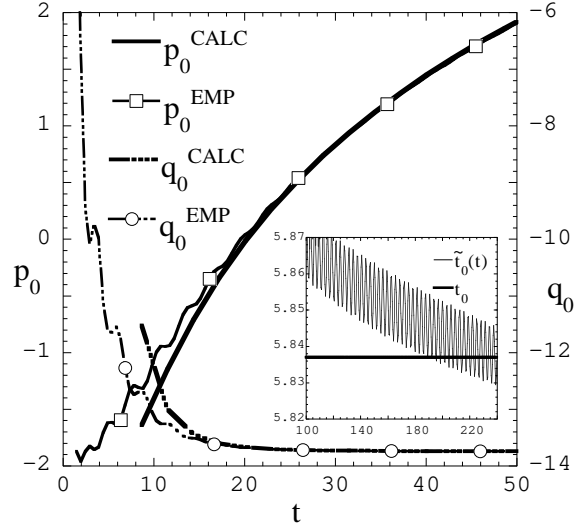}
\caption{\label{fig7} An alternative approach to the determination of $t_0$. The
empirically determined $\tilde p_0$ (labeled $p_0^{EMP}$) and $\tilde q_0$
approach $p_0$ (labeled $p_0^{CALC}$) and $q_0$ as $\tilde t_0$ evolves toward $t_0$
(inset).}
\end{figure}

\begin{figure}
\includegraphics[scale = .5]{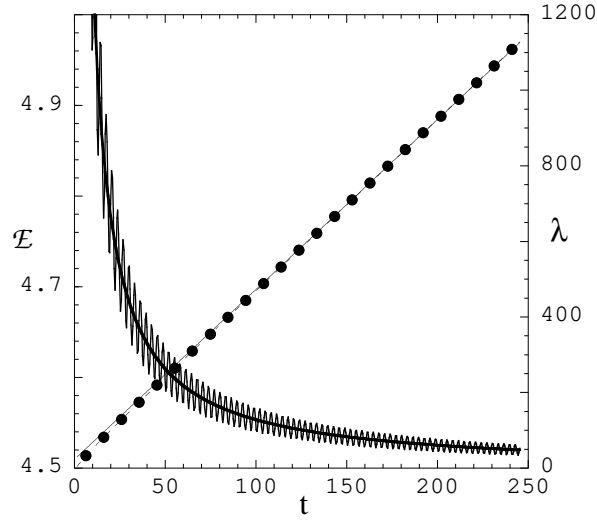}
\caption{\label{fig8} Linear growth of $\lambda$. The dots represent every 20th
point in the spatial average of $\lambda$ as computed using Eq.~(\ref{dldt}) in a
numerical simulation. The solid line is $\bar \lambda = 23.00 \,+\,4.4969\,t$. Also
shown is the energy-like spatial average of the source terms for $\lambda,_t$,
which, while fluctuating, asymptote to a constant. Note the small range on the scale
for ${\cal E}$. The thick solid line is ${\cal E} = 4.4969 \,+\, 5.6388/t$.}
\end{figure}

\begin{figure}
\includegraphics[scale = .5]{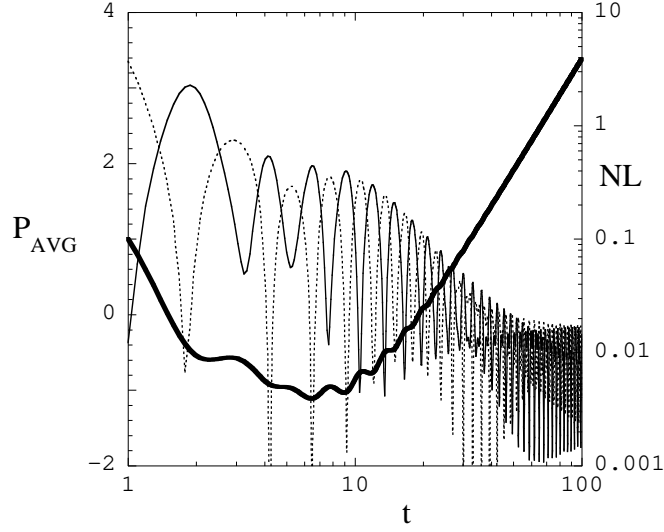}
\caption{\label{fig9} Nonlinear fluctuations and the spatial average of $P$. The
nonlinear terms $V_1 = \oint d\theta \,e^{2P}Q,_t^2$ (thin solid line) and $V_2 =
\oint d\theta \, e^{2P}Q,_\theta^2$ (dashed line) are shown on a log-log plot (using
the vertical axis labeled $NL$). The logarithmic time axis is used to emphasize the
behavior during the early stages of the simulation when nonlinear effects are
significant. For comparison, the spatial average of $P$ is also shown (thick solid
line). Note that $V_1$ increases and $V_2$ decreases as $P$ decreases as expected
if these terms are acting to regulate $P$.}
\end{figure}

\begin{figure}
\includegraphics[scale = .5]{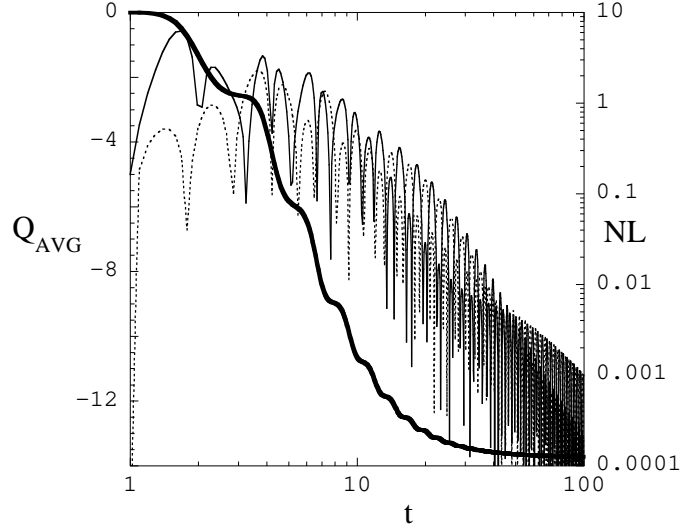}
\caption{\label{fig10}  Nonlinear fluctuations and the spatial average of $Q$. The
nonlinear terms $W_1 = 2| \oint d\theta \, P,_tQ,_t|$ (thin solid line) and $W_2 = 2
|\oint d\theta \,P,_\theta Q,_\theta|$ (dashed line) are shown on a log-log plot
(using the vertical axis labeled $NL$). For comparison, the spatial average of $Q$
is also shown (thick solid line).}
\end{figure}

\begin{figure}
\includegraphics[scale = .5]{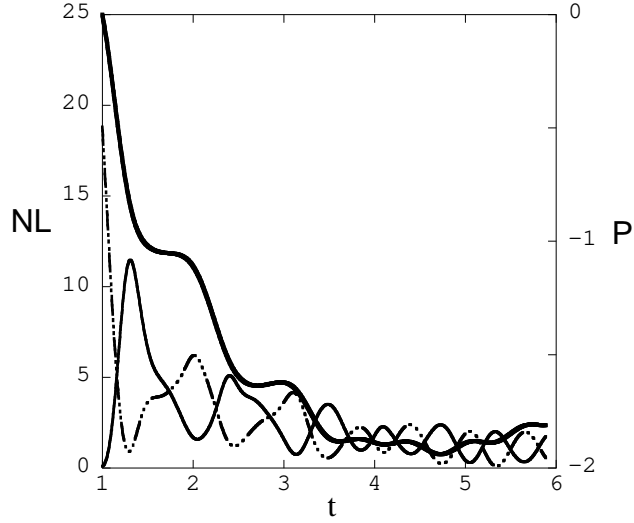}
\caption{\label{fig12}  Nonlinear terms and the spatial average of $P$ early in the
simulation. The nonlinear terms $V_1$ (thin solid line) and $V_2$ (dashed line)
(see Fig.~\ref{fig9}) are shown for $1 \le t \le 6$ (using the vertical axis labeled
$NL$). For comparison, the spatial average of $P$ is also shown (thick solid line).
For this and subsequent figures, the initial data are $P = 0$, $P,_t = 2\,+\, 2 \cos
\theta
\,+\, 0.5 \cos (2\theta) \,+\, 4 \cos (5\theta)$, $Q = \cos \theta \,+\, 2 \cos
(3\theta)
\,+\, \cos (7\theta)$, $Q,_t = 0.3$. Note that, as $P$ becomes increasingly
negative early in the simulation, $V_1$ increases while $V_2$ decreases supporting
the contention that the former evolves as $e^{-2P}$. Also note the qualitative
similarities to Fig.~\ref{fig9} despite quite different initial data.} 
\end{figure}

\begin{figure}
\includegraphics[scale = .5]{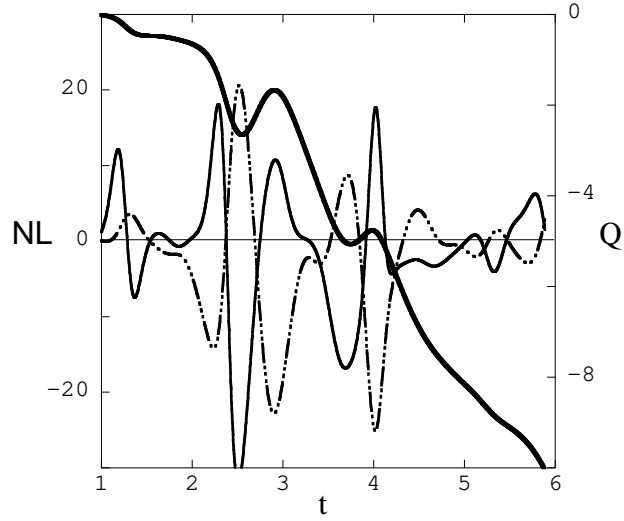}
\caption{\label{fig13}  Nonlinear terms and the spatial average of $Q$ early
in the simulation. The nonlinear terms $W_1$ (thin solid line) and $W_2$ (dashed
line) (see Fig.~\ref{fig10}) are shown for $1 \le t \le 6$ (using the vertical axis
labeled $NL$). For comparison, the spatial average of $Q$ is also shown (thick
solid line). The horizontal line indicates zero for the nonlinear terms. Note that
the difference in initial data leads to qualitative differences in evolution in
Figs.~\ref{fig10} and \ref{fig13}.}
\end{figure}

\begin{figure}
\includegraphics[scale = .5]{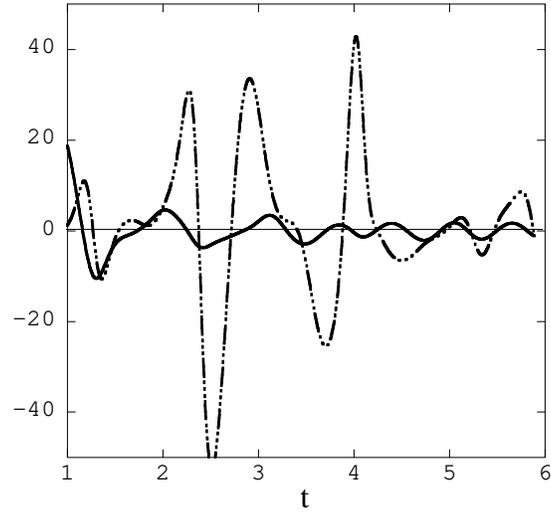}
\caption{\label{fig14}  Nonlinear source terms $S_P(t)$ (solid line) and $S_Q(t)$
(dashed line) early in the simulation. Note how quickly these terms approach a
state where they oscillate more or less symmetrically about zero.}
\end{figure}

\begin{figure}
\includegraphics[scale = .5]{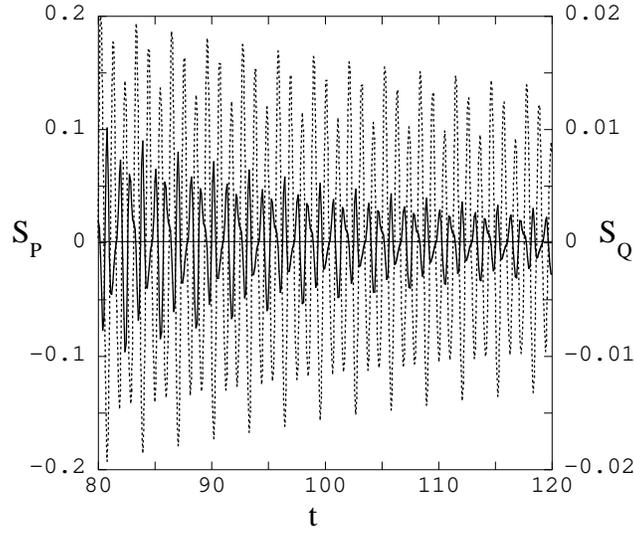}
\caption{\label{fig15}  Nonlinear source terms $S_P(t) = V_1 \,-\, V_2$ (solid line)
and
$S_Q(t) = W_2 \,-\, W_1$ (dashed line) late in the simulation. Note the small
amplitude of these terms and that their time averages are very close to zero ($9
\times 10^{-4}$ and
$2 \times 10^{-5}$ respectively).}
\end{figure}


\begin{thebibliography}{24}
\expandafter\ifx\csname natexlab\endcsname\relax\def\natexlab#1{#1}\fi
\expandafter\ifx\csname bibnamefont\endcsname\relax
  \def\bibnamefont#1{#1}\fi
\expandafter\ifx\csname bibfnamefont\endcsname\relax
  \def\bibfnamefont#1{#1}\fi
\expandafter\ifx\csname citenamefont\endcsname\relax
  \def\citenamefont#1{#1}\fi
\expandafter\ifx\csname url\endcsname\relax
  \def\url#1{\texttt{#1}}\fi
\expandafter\ifx\csname urlprefix\endcsname\relax\def\urlprefix{URL }\fi
\providecommand{\bibinfo}[2]{#2}
\providecommand{\eprint}[2][]{\url{#2}}

\bibitem[{\citenamefont{Kasai}(1993)}]{kasai93}
\bibinfo{author}{\bibfnamefont{M.}~\bibnamefont{Kasai}},
  \bibinfo{journal}{Phys.\ Rev.\ D} \textbf{\bibinfo{volume}{47}},
  \bibinfo{pages}{3214} (\bibinfo{year}{1993}).

\bibitem[{\citenamefont{Tomita and Deruelle}(1994)}]{tomita94}
\bibinfo{author}{\bibfnamefont{K.}~\bibnamefont{Tomita}} \bibnamefont{and}
  \bibinfo{author}{\bibfnamefont{N.}~\bibnamefont{Deruelle}},
  \bibinfo{journal}{Phys.\ Rev.\ D} \textbf{\bibinfo{volume}{50}},
  \bibinfo{pages}{7216} (\bibinfo{year}{1994}).

\bibitem[{\citenamefont{Deruelle and Langlois}(1995)}]{deruelle95}
\bibinfo{author}{\bibfnamefont{N.}~\bibnamefont{Deruelle}} \bibnamefont{and}
  \bibinfo{author}{\bibfnamefont{D.}~\bibnamefont{Langlois}},
  \bibinfo{journal}{Phys.\ Rev.\ D} \textbf{\bibinfo{volume}{52}},
  \bibinfo{pages}{2007} (\bibinfo{year}{1995}).

\bibitem[{\citenamefont{Carfora and Piotrkowska}(1995)}]{carfora95}
\bibinfo{author}{\bibfnamefont{M.}~\bibnamefont{Carfora}} \bibnamefont{and}
  \bibinfo{author}{\bibfnamefont{K.}~\bibnamefont{Piotrkowska}},
  \bibinfo{journal}{Phys.\ Rev.\ D} \textbf{\bibinfo{volume}{52}},
  \bibinfo{pages}{4393} (\bibinfo{year}{1995}).

\bibitem[{\citenamefont{Futamase}(1996)}]{futamase96}
\bibinfo{author}{\bibfnamefont{T.}~\bibnamefont{Futamase}},
  \bibinfo{journal}{Phys.\ Rev.\ D} \textbf{\bibinfo{volume}{53}},
  \bibinfo{pages}{681} (\bibinfo{year}{1996}).

\bibitem[{\citenamefont{Stoeger et~al.}()\citenamefont{Stoeger, Helmi, and
  Torres}}]{stoeger99}
\bibinfo{author}{\bibfnamefont{W.~R.} \bibnamefont{Stoeger}},
  \bibinfo{author}{\bibfnamefont{A.}~\bibnamefont{Helmi}}, \bibnamefont{and}
  \bibinfo{author}{\bibfnamefont{D.~F.} \bibnamefont{Torres}},
  \emph{\bibinfo{title}{Averaging {E}instein's equations: The linearized
  case}}, \eprint{gr-qc/9904020}.

\bibitem[{\citenamefont{Buchert}()}]{buchert00}
\bibinfo{author}{\bibfnamefont{T.}~\bibnamefont{Buchert}},
  \emph{\bibinfo{title}{On average properties of inhomogeneous cosmologies}},
  \eprint{gr-qc/0001056}.

\bibitem[{\citenamefont{Gowdy}(1971)}]{gowdy71}
\bibinfo{author}{\bibfnamefont{R.~H.} \bibnamefont{Gowdy}},
  \bibinfo{journal}{Phys.\ Rev.\ Lett.} \textbf{\bibinfo{volume}{27}},
  \bibinfo{pages}{826} (\bibinfo{year}{1971}).

\bibitem[{\citenamefont{Berger}(1974)}]{berger74}
\bibinfo{author}{\bibfnamefont{B.~K.} \bibnamefont{Berger}},
  \bibinfo{journal}{Ann.\ Phys.\ (N.Y.)} \textbf{\bibinfo{volume}{83}},
  \bibinfo{pages}{458} (\bibinfo{year}{1974}).

\bibitem[{\citenamefont{Belinskii et~al.}(1971)\citenamefont{Belinskii,
  Lifshitz, and Khalatnikov}}]{belinskii71b}
\bibinfo{author}{\bibfnamefont{V.~A.} \bibnamefont{Belinskii}},
  \bibinfo{author}{\bibfnamefont{E.~M.} \bibnamefont{Lifshitz}},
  \bibnamefont{and} \bibinfo{author}{\bibfnamefont{I.~M.}
  \bibnamefont{Khalatnikov}}, \bibinfo{journal}{Sov.\ Phys.\ Usp.}
  \textbf{\bibinfo{volume}{13}}, \bibinfo{pages}{745} (\bibinfo{year}{1971}).

\bibitem[{\citenamefont{Isenberg and Moncrief}(1990)}]{isenberg90}
\bibinfo{author}{\bibfnamefont{J.~A.} \bibnamefont{Isenberg}} \bibnamefont{and}
  \bibinfo{author}{\bibfnamefont{V.}~\bibnamefont{Moncrief}},
  \bibinfo{journal}{Ann.\ Phys.\ (N.Y.)} \textbf{\bibinfo{volume}{199}},
  \bibinfo{pages}{84} (\bibinfo{year}{1990}).

\bibitem[{\citenamefont{Berger and Moncrief}(1993)}]{berger93}
\bibinfo{author}{\bibfnamefont{B.~K.} \bibnamefont{Berger}} \bibnamefont{and}
  \bibinfo{author}{\bibfnamefont{V.}~\bibnamefont{Moncrief}},
  \bibinfo{journal}{Phys.\ Rev.\ D} \textbf{\bibinfo{volume}{48}},
  \bibinfo{pages}{4676} (\bibinfo{year}{1993}).

\bibitem[{\citenamefont{Berger and Garfinkle}(1998)}]{berger97b}
\bibinfo{author}{\bibfnamefont{B.~K.} \bibnamefont{Berger}} \bibnamefont{and}
  \bibinfo{author}{\bibfnamefont{D.}~\bibnamefont{Garfinkle}},
  \bibinfo{journal}{Phys.\ Rev.\ D} \textbf{\bibinfo{volume}{57}},
  \bibinfo{pages}{4767} (\bibinfo{year}{1998}).

\bibitem[{\citenamefont{Centrella and Matzner}(1982)}]{centrella82}
\bibinfo{author}{\bibfnamefont{J.}~\bibnamefont{Centrella}} \bibnamefont{and}
  \bibinfo{author}{\bibfnamefont{R.~A.} \bibnamefont{Matzner}},
  \bibinfo{journal}{Phys.\ Rev.\ D} \textbf{\bibinfo{volume}{25}},
  \bibinfo{pages}{930} (\bibinfo{year}{1982}).

\bibitem[{\citenamefont{Anninos}(1998)}]{anninos98}
\bibinfo{author}{\bibfnamefont{P.}~\bibnamefont{Anninos}},
  \bibinfo{journal}{Phys.\ Rev.\ D} \textbf{\bibinfo{volume}{58}},
  \bibinfo{pages}{064010} (\bibinfo{year}{1998}).

\bibitem[{\citenamefont{Anninos and McKinney}(1999)}]{anninos99}
\bibinfo{author}{\bibfnamefont{P.}~\bibnamefont{Anninos}} \bibnamefont{and}
  \bibinfo{author}{\bibfnamefont{J.}~\bibnamefont{McKinney}},
  \bibinfo{journal}{Phys.\ Rev.\ D} \textbf{\bibinfo{volume}{60}},
  \bibinfo{pages}{064011} (\bibinfo{year}{1999}).

\bibitem[{\citenamefont{Berger et~al.}(1998)\citenamefont{Berger, Garfinkle,
  and Moncrief}}]{berger97c}
\bibinfo{author}{\bibfnamefont{B.~K.} \bibnamefont{Berger}},
  \bibinfo{author}{\bibfnamefont{D.}~\bibnamefont{Garfinkle}},
  \bibnamefont{and} \bibinfo{author}{\bibfnamefont{V.}~\bibnamefont{Moncrief}},
  in \emph{\bibinfo{booktitle}{Internal Structure of Black Holes and Spacetime
  Singularities}}, edited by
  \bibinfo{editor}{\bibfnamefont{L.}~\bibnamefont{Burko}} \bibnamefont{and}
  \bibinfo{editor}{\bibfnamefont{A.}~\bibnamefont{Ori}}
  (\bibinfo{publisher}{Institute of Physics}, \bibinfo{address}{Bristol},
  \bibinfo{year}{1998}), p.~\bibinfo{pages}{44}.

\bibitem[{\citenamefont{Berger and Moncreif}(2000)}]{berger00}
\bibinfo{author}{\bibfnamefont{B.~K.} \bibnamefont{Berger}} \bibnamefont{and}
  \bibinfo{author}{\bibfnamefont{V.}~\bibnamefont{Moncreif}},
  \bibinfo{journal}{Phys.\ Rev.\ D} \textbf{\bibinfo{volume}{62}},
  \bibinfo{pages}{023509} (\bibinfo{year}{2000}).

\bibitem[{\citenamefont{Abramowitz and Stegun}(1965)}]{abramowitz65}
\bibinfo{author}{\bibfnamefont{M.}~\bibnamefont{Abramowitz}} \bibnamefont{and}
  \bibinfo{author}{\bibfnamefont{I.~A.} \bibnamefont{Stegun}},
  \emph{\bibinfo{title}{Handbook of Mathematical Functions with Formulas,
  Graphs, and Mathematical Tables}} (\bibinfo{publisher}{Dover},
  \bibinfo{address}{New York}, \bibinfo{year}{1965}).

\bibitem[{\citenamefont{Arfken and Weber}(2000)}]{arfken00}
\bibinfo{author}{\bibfnamefont{G.~B.} \bibnamefont{Arfken}} \bibnamefont{and}
  \bibinfo{author}{\bibfnamefont{H.-J.} \bibnamefont{Weber}},
  \emph{\bibinfo{title}{Mathematical Methods for Physicists}}
  (\bibinfo{publisher}{Harcourt/Academic}, \bibinfo{address}{San Diego},
  \bibinfo{year}{2000}).

\bibitem[{\citenamefont{Garfinkle}(1999)}]{garfinkle99}
\bibinfo{author}{\bibfnamefont{D.}~\bibnamefont{Garfinkle}},
  \bibinfo{journal}{Phys.\ Rev.\ D} \textbf{\bibinfo{volume}{60}},
  \bibinfo{pages}{104010} (\bibinfo{year}{1999}).

\bibitem[{\citenamefont{Ringstr{\"{o}}m}()}]{ringstrom01}
\bibinfo{author}{\bibfnamefont{H.}~\bibnamefont{Ringstr{\"{o}}m}},
  \bibinfo{note}{private communication}.

\bibitem[{\citenamefont{Berger et~al.}(1997)\citenamefont{Berger, Isenberg,
  Chru\'{s}ciel, and Moncrief}}]{berger97g}
\bibinfo{author}{\bibfnamefont{B.~K.} \bibnamefont{Berger}},
  \bibinfo{author}{\bibfnamefont{J.}~\bibnamefont{Isenberg}},
  \bibinfo{author}{\bibfnamefont{P.~T.} \bibnamefont{Chru\'{s}ciel}},
  \bibnamefont{and} \bibinfo{author}{\bibfnamefont{V.}~\bibnamefont{Moncrief}},
  \bibinfo{journal}{Ann.\ Phys.} \textbf{\bibinfo{volume}{260}},
  \bibinfo{pages}{117} (\bibinfo{year}{1997}).

\bibitem[{\citenamefont{Berger and Isenberg}()}]{berger01}
\bibinfo{author}{\bibfnamefont{B.~K.} \bibnamefont{Berger}} \bibnamefont{and}
  \bibinfo{author}{\bibfnamefont{J.}~\bibnamefont{Isenberg}},
  \emph{\bibinfo{title}{Asymptotic behavior of expanding ${T}^2$ symmetric
  spacetimes}}, \bibinfo{note}{unpublished}.

\end{thebibliography}
\end{document}